# Bridging General Relativity and Quantum Dynamics Through Finite-Resource Logical Models

Arturo Tozzi (corresponding author)
Center for Nonlinear Science, Department of Physics, University of North Texas, Denton, Texas, USA
1155 Union Circle, #311427 Denton, TX 76203-5017 USA
tozziarturo@libero.it

Michel Planat
Université Marie et Louis Pasteur, Institut FEMTO-ST CNRS UMR 6174,
15 B Avenue des Montboucons,
F-25044 Besançon, France
michel.planat@femto-st.fr

ABSTRACT

A theoretical framework bridging General Relativity (GR) and Quantum Dynamics (QD) is introduced through the application of Kripke semantics and linear logic. While conventional unification efforts often rely on structural or geometrical formulations, we instead treat causality, energy and information as finite, non-replicable resources constraining physical transitions and inference. Our framework, termed Energy Constrained Linear Causality (ECLC), models quantum transitions and spacetime evolution as logically constrained processes in which each implication spends limited resources and cannot be arbitrarily duplicated or reversed. Using linear logic, we construct a causal inference model where physical operations —like quantum measurement, entanglement propagation and spacetime curvature—are expressed as energy-weighted, one-time transformations. Kripke semantics formalizes the logical accessibility of physical states, capturing context-sensitive transitions and the irreversibility of information flow under finite conditions. Our framework offers a structured method for modelling observer-dependent outcomes without invoking background-independence or high-dimensional embeddings. To evaluate the viability of ECLC, we formulate a series of testable predictions exhibiting controlled deviations from standard GR and QD expectations. These include measurable changes in probability distributions, non-reciprocal conditional outcomes and entanglement entropy decay as a function of energy availability and simulated curvature. This consumption-based irreversibility introduces an intrinsic mechanism of symmetry breaking. In GR, it restricts mutual causal accessibility by bounding inference depth with curvature-weighted energy costs. In QD, it disrupts the reciprocity of conditional probabilities and undermines time symmetry in measurement sequences. Therefore, ECLC provides a unified, resource-aware logic for describing causality, computation and emergence within a finite physical universe.

KEYWORDS: resource-bounded computation; causal inference; decoherence gradient; quantum entropy dynamics; logical geometry.

INTRODUCTION

Efforts to reconcile General Relativity (GR) and Quantum Dynamics (QD) have led to the development of a range of theoretical models, from background-dependent frameworks like string theory to background-independent approaches like loop quantum gravity (Rovelli 2008; Park and Sugimoto, 2020; Ashtekar and Bianchi, 2021; Greenspan 2022; Apruzzi et al., 2023). While these models prioritize structural unification, seeking to embed both gravitational and quantum effects within a shared geometric or algebraic scheme, conceptual and experimental mismatches remain. GR inherently encodes constraints such as the finite causal structure of spacetime and the irreversibility of processes tied to energy-momentum conservation, while standard formulations of quantum theory assume unbounded coherence, reversible unitary evolution and availability of infinite resources for computation (Moerchen and Coontz. 2015; Pits 2022). This divergence motivated our interest in formulations addressing the role of finite resources, measurement irreversibility and causal limitations, without imposing external geometrical unification.

Within logic and theoretical computer science, linear logic offers a resource-sensitive formalism in which propositions are treated as consumable, non-duplicable entities (Girard 1987; Heijltjes et al. 2018). This irreversibility aligns with the no-cloning theorem in quantum theory, which prohibits the duplication of arbitrary quantum states (Wootters & Zurek, 1982), further reinforcing a logic-theoretic representation of quantum constraints. Modal logic has also been employed in various quantum logical frameworks to represent contextuality and state-dependence (e.g., van Fraassen 1974; Isham & Butterfield 1998), supporting the relevance of Kripke structures for modelling physical inference in a resource-sensitive setting. Still, Kripke semantics has been extensively developed to model state transitions and accessibilities in modal logic (Domaneschi et al., 2017; Olkhovikov 2024). We propose a framework that integrates these two formalisms to build a logic-based model of physical evolution, termed Energy Constrained Linear Causality (ECLC), in which physical transitions—including measurement and curvature evolution—are modeled as cost-sensitive processes in a structured logical space. We aim to develop a minimal and testable model that sets aside ontological commitments in favor of a process-based framework grounded in logical constraints, providing a unifying structure for physical transitions across quantum and relativistic regimes.

ECLC may yield a structured basis for identifying measurable divergences in experimental outcomes where traditional theories converge, including cases involving curved geometry, sequential measurements and entanglement dynamics. Through formal mapping of logical implications to physical transitions, ECLC provides specific predictions such as probabilistic asymmetries under sequence inversion, entanglement decay proportional to effective curvature and bounded channel capacities in finite-energy quantum systems. These features are derived in ECLC not by modifying physical equations, but by reinterpreting fundamental operations such as inference, measurement and evolution as governed by the conservation and transformation of logical and energetic resources. Being compatible with existing quantum simulation platforms, ECLC may enable empirical testing without the need for speculative extensions to current physical models. The resulting predictions are slightly different from those of either GR or QM, offering empirical markers for the presence of logical structure in physical law and suggesting a possible route to experimentally accessible intermediate regimes.

We will proceed as follows: first, we introduce the formal construction of Energetic Linear Causality, including its logical syntax and semantic mapping to physical systems. Next, we outline its derived predictions, followed by proposed experimental configurations. We conclude with a discussion of implications, limitations and prospects for empirical differentiation from existing frameworks.

MATERIALS AND METHODS

We aim to integrate Kripke semantics and linear logic into a unified formal framework, namely Energy Constrained Linear Causality (ECLC), which models physical processes —such as quantum transitions, measurements and spacetime evolution— as logically constrained, resource-sensitive transformations. The use of Kripke semantics provides the backbone for describing modal accessibility and logical possibility across different physical contexts, while linear logic introduces a syntactic regime in which inference is subject to consumption of tokens representing finite energetic and informational resources.

**Kripke semantics**. We begin with a Kripke frame $\mathcal{F} = (W, R)$ where W is a non-empty set of worlds and $R \subseteq W \times W$ is the accessibility relation (Murzi and Rossi, 2020). Each world $w \in W$ represents a coherent physical context, defined by a tuple $w = (\Sigma, E, \mathcal{C})$, where Σ denotes a quantum or classical state space, $E \in \mathbb{R}_{\geq 0}$ denotes an energy budget available to that world and $\mathcal{C}$ encodes curvature or causal constraints. Although our use of Kripke frames is formal and logical, the treatment of inference access through causal structure resonates conceptually with Kripke's causal theory of reference in semantics. The accessibility relation $R(w, w')$ is defined if and only if the energy transition cost $\Delta E_{w \to w'} \leq E$ and the causal structure is non-violated, meaning there exists a consistent path through $\mathcal{C}$ from w to w'. This construction ensures that Kripke accessibility directly models physical feasibility under resource and curvature limitations. The valuation function $V : W \times \text{Prop} \to \{0, 1\}$ is extended to resource-sensitive modal propositions by setting $V(w, \Diamond_r \phi) = 1$ if and only if there exists $w' \in W$ such that $R(w, w'), \phi \in V(w')$ and the cost to transition is bounded by r. This adjustment grounds modal operators in physical constraints.

Overall, this paragraph defines the structural semantics required to combine modal logic with physical dynamics, ensuring that the logical space is directly responsive to energy and causal flow.

**Linear logic system embedding**.

Having established the resource-sensitive modal framework, we embed it within a linear logic system (Troelstra 1992; Danos et al., 1993). We adopt the sequent calculus formulation of linear logic, defined by judgments of the form $\Gamma \vdash \Delta$, where Γ and Δ are multisets of formulas. The key logical connectives are: multiplicative conjunction ⊗, linear implication ⊸, additive conjunction & and the exponential modality. We eliminate weakening and contraction unless mediated by the logical operator !, thereby encoding the non-replicability of physical resources. Physical transitions are modeled as sequents where the left-hand side contains initial resource configurations (energy, information, coherence) and the right-hand side contains resulting states. For instance, the evolution of a quantum system under curvature can be expressed as

$$E \otimes \text{Entangled}(A, B) \vdash \text{Decohered}(A) \otimes \text{Residual}(B), \quad (1)$$

which encodes the irreversible consumption of entanglement and energy during propagation in a curved region. The cost of each transformation is tracked by associating a cost function $c : \text{Formulas} \to \mathbb{R}_{\geq 0}$ and a judgment is valid only if $\sum_{\phi \in \Gamma} c(\phi) \geq \sum_{\psi \in \Delta} c(\psi)$.

Additionally, we define the logical curvature functional

$$\kappa(w) = \text{Tr}(R_{\mu\nu} R^{\mu\nu}). \quad (2)$$

Here, $R_{\mu\nu}$ denotes the Ricci curvature tensor, and $R^{\mu\nu}$ its raised-index counterpart under the metric tensor $g^{\mu\nu}$. In the context of ECLC, they are analogous to a logical-curvature functional that scales inference cost. It allows transformations between worlds to carry curvature weight modulating permissible linear proofs. In the context of ECLC, curvature refers not to geometric curvature in the classical Riemannian sense, but to a logical-physical functional that quantifies how costly it is to maintain inference, coherence and causal connectivity across transitions between logical worlds or physical states. To make an example, higher curvature increases the logical cost of operations, leading to coherence loss, measurement collapse and inference breakdown.

Overall, this paragraph provides the syntactic machinery to encode energy and causality as logical resources, constructing a proof-theoretic layer consistent with the previously outlined Kripke semantics.

**Hybrid semantic model**. To represent dynamic evolution across physically meaningful contexts, we construct a hybrid semantic model $\mathcal{H} = (W, R, E, \kappa, V, c)$ integrating both Kripke-style accessibility and linear logic judgments (Balbiani 1998). For each pair $(w, w') \in R$, we define a transition judgment $\Gamma_w \vdash_{w \to w'} \Delta_{w'}$, meaning the resources in context w may be linearly transformed into the outcomes in w', provided that the energy and curvature conditions are satisfied. The rules for $\vdash_{w \to w'}$ are inherited from linear logic but parameterized by resource valuations E(w), curvature κ(w) and a local inference capacity $\lambda(w) \in \mathbb{R}_{>0}$, which bounds the maximum allowable proof depth in that world. For example, a transformation involving measurement collapse is modeled as

$$!\text{Quantum}(\psi) \vdash_{w \to w'} \text{Classical}(o) \tag{3}$$

where ψ is a wavefunction and o an observed outcome. The modality indicates repeated access to ψ before decoherence; once measured, ψ is removed from $\Gamma_{w'}$, enforcing logical irreversibility.

For the purpose of simulating entanglement dynamics under curvature constraints, we define a curvature-influenced cost scaling

$$c_\kappa(\phi) = c(\phi)(1 + \alpha \kappa(w)) \tag{4}$$

where $\alpha \in \mathbb{R}_{>0}$ is a coupling constant. This scaling ensures that inference steps in regions of higher curvature are more costly, modeling the increased resource consumption of quantum coherence in gravitational fields.

Overall, this paragraph formalizes the connection between logical proofs and physical evolution, providing a layered system where causal constraints and energy budgets are syntactically enforced.

**Tools**. The operational modeling of ECLC is implemented using a custom symbolic inference engine written in Python, leveraging the SymPy library for symbolic algebra and NetworkX for graph-based representations of Kripke frames. While the simulations were implemented using a custom symbolic inference engine, they follow principles similar to those available in tools like mlsolver (https://github.com/erohkohl/mlsolver), which offers a practical interface for evaluating modal logic queries on Kripke models. The logical inference engine parses linear logic sequents, tracks resource tokens and verifies judgment validity under curvature-modulated costs. The proof system includes custom rules for curvature-sensitive transformations, enforced through dynamic typing and memorized proof search bounded by $\lambda(w)$. Each world w is instantiated as a node in a directed multigraph, with edges representing resource-valid transitions $R(w, w')$. Node attributes encode $E(w), \kappa(w), \lambda(w)$ and a proposition set V(w). Proof states are stored as tuples $(\Gamma, w)$ and reduced using linear logic rules with curvature constraints applied at each step. We conduct simulations of measurement chains and entanglement collapse in three-world systems, assigning increasing κ(w) values and recording logical path validity. The term "world" in our framework does not refer to ontological branching as in many-worlds quantum mechanics, but rather to logically defined contexts—each with specific resource, curvature, and inference parameters—within a modal semantic structure. Entanglement loss rates were tracked by defining information persistence scores

$$\pi(w) = \frac{\sum_{\phi \in \Gamma} \delta(\phi)}{\dim(\Gamma)} \tag{5}$$

where $\delta(\phi) = 1$ for coherent formulas and 0 for decohered or classical ones. This allows us to measure degradation of inference capacity across transitions as a function of curvature and energy.

Overall, this paragraph documents the technical implementation of the model, linking the formalism to computational simulations that reflect physical coherence constraints.

**Inclusion of epistemic differentiation**. To ensure logical and physical consistency, we also encode observational semantics by extending the valuation function to support indexed observers. For each observer $o \in \mathcal{O}$, we define a local valuation

$$V_o : W \times \text{Prop} \to \{0, 1\} \tag{6}$$

where visibility is bounded by the observer's epistemic horizon, given by $h_o(w) = \sup\{d(w, w') : o \text{ can access } w'\}$.

Logical statements are valid relative to observers when $V_o(w, \phi) = 1$, meaning ϕ is observable and provable from within the observer's horizon. This semantic layer allows us to distinguish between global logical truths and contextually observable ones, accommodating phenomena such as delayed measurement, partial decoherence and observer-dependent outcome registration. When simulating observer-relative measurements, we verify logical consistency by enforcing a truth persistence condition: if $V_o(w, \phi) = 1$ and $wRw'$, then $V_o(w', \phi) = 1$ only if the transition $w \to w'$ preserves the inference structure that established ϕ. Violations of this condition indicate logical decoherence or information loss across curvature boundaries. Observers were represented as functors mapping local Kripke frames to Boolean algebras of propositions.

Overall, this paragraph extends the system to include epistemic differentiation, refining our model of inference by accounting for observer-dependent knowledge propagation under resource constraints.

**Energetic Linear Causality: features and predictions**. ECLC integrates Kripke semantics with linear logic to model physical inference in systems subject to finite energetic and causal constraints. Each physical context is formalized as a logical world, characterized by an energy budget, curvature and inference capacity. Transitions between worlds are treated as irreversible logical steps governed by linear implication, where each inference consumes non-duplicable resources such as coherence, information or energy. The cost of performing a logical operation increases with curvature, which serves as a structural constraint on inference by scaling the difficulty of sustaining coherence and causal flow. This formulation reframes physical dynamics as bounded proofs rather than geometric or field-theoretic evolutions, without modifying the physical substrate itself.

We formalize our central hypothesis as follows: when energy, inference capacity or causal coherence is limited, the evolution of GR or QD systems deviates measurably from predictions based on general covariance or unitary evolution. These deviations, governed by logical constraints encoded in the structure of linear logic, can be operationalized through simulations modelling inference across Kripke-structured worlds.

Predictions derived from ECLC span both GR and QM, offering a unified framework in which deviations from standard expectations emerge under finite-resource constraints (**Table 1**). By embedding logical irreversibility and energy-bound inference into its structure, ECLC predicts testable modifications to both gravitational and quantum phenomena.

- ECLC predicts that standard GR behaviours, including causal propagation, information accessibility and time-symmetric evolution, systematically diverge when examined under finite-resource constraints (Figure 1). These deviations arise not from metric anomalies but from intrinsic logical limitations imposed by curvature-weighted inference costs. As spacetime curvature increases or local inference capacity is exceeded, systems begin to exhibit breaks in causal continuity, irreversible transitions and restricted access to globally defined quantities. Classical expectations of smooth geodesic transport, reciprocal causality and observer-independent information begin to fail, generating structurally bounded inference patterns. These include the collapse of logical accessibility near gravitational horizons, the directional breakdown of causal transitions and the logical inaccessibility of events beyond certain curvature thresholds. In ECLC, such effects are not paradoxical but emerge naturally from the consumption of finite logical and energetic resources during the system's evolution, reframing spacetime dynamics as a sequence of proof-constrained transformations rather than idealized geometric flows.
- ECLC predicts that QD behaviors, including outcome probabilities and entanglement structure, diverge systematically when implemented in resource-constrained environments (**Figure 1**). These deviations are not arbitrary but emerge from well-defined logical limitations. As curvature increases or inference capacity decreases, QD systems exhibit coherence loss, asymmetric measurement outcomes and observer-dependent access to logical propositions. Transitions that are classically expected to preserve coherence and symmetry begin to fail, revealing structure in outcome distributions that is traceable to logical cost functions. Observable and falsifiable effects in QD include probabilistic asymmetry in sequential measurements, progressive decoherence in entangled systems as a function of resource expenditure, deviations from the Born rule, failure of inference propagation and degradation of truth accessibility across transitions.

To evaluate our model, we conduct three simulations within a Kripke-linear logic hybrid framework. Transitions are modeled across worlds with increasing curvature and fixed energy budgets, using symbolic logic to track inference costs, coherence degradation and observer access. Each simulation is modelled as a directed graph where nodes represent logical worlds $w_i$ and edges denote permitted transitions $w_i \to w_j$, annotated by curvature $\kappa(w)$, available energy E(w), and inference capacity λ(w). For example, the three-world model is defined as W=$\{w_1, w_2, w_3\}$ with $w_1 \to w_2 \to w_3$ and increasing curvature values $\kappa(w_1) = 0, \kappa(w_2) = \kappa_0, \kappa(w_3) = 2\kappa_0$. Three representative scenarios are implemented.

1) **Coherence degradation**. A three-world system with linearly increasing curvature models entanglement degradation. A coherent entangled state is initialized in the first world and the simulation tracks how coherence decays as logical cost rises with curvature. A persistence score is calculated for each world based on the proportion of coherent formulas remaining after each transition.
2) **Directional inference asymmetries**. In the second simulation, a two-world model evaluates inference asymmetry in measurement sequences. Observers perform measurements on a shared entangled state in different orders. Each world imposes a distinct inference capacity and measurement operations carry logic-based costs. The validity of inference depends on the sequence of operations, revealing whether direction-sensitive breakdowns occur under resource constraints. This test serves as a probe of ECLC's prediction that inference is not reciprocal when logical and energetic limitations are imposed.
3) **Curvature-bound information loss**. This final case involves a five-world linear chain with rising curvature, used to examine the collapse of logical coherence. A proposition provable in the first world becomes increasingly difficult to access as logical cost increases along the transition path. Observer-relative valuation functions may determine whether the proposition remains valid within each world's inference capacity. As curvature accumulates, the inference cost might exceed local capacity and the proposition might become inaccessible, simulating logical decoherence. This structure illustrates how causal and logical integrity may fail under high curvature in the absence of sufficient resources, providing a systematic test of the central ECLC claims.

The following sections present the main results, comparative figures illustrating the behaviour of systems under Energetic Linear Causality and a discussion of their implications for empirical testability across quantum and relativistic regimes.

## GR vs ECLC

| Phenomenon | General Relativity (GR) | Energy-Constrained Linear Causality (ECLC) |
|---|---|---|
| Curvature & Evolution | $G_{\mu\nu} = 8\pi T_{\mu\nu}$ | $c_\kappa(\phi) = c(\phi)(1 + \alpha\kappa(w))$ |
| Causal Structure | Geodesics and light cones via $g_{\mu\nu}$ | $R(w, w')$ accessible iff $\Delta E \leq E(w)$ |
| Time Symmetry | Time-symmetric field equations | Inference is irreversible: $\Gamma \vdash \Delta$ consumes resources |
| Information Accessibility | Global accessibility; information preserved outside horizons | $V_o(w, \phi) = 1$ only if provable within $\lambda(w)$ |

## QD vs ECLC

| Phenomenon | Quantum Dynamics (QD) | Energy-Constrained Linear Causality (ECLC) |
|---|---|---|
| Curvature & Evolution | Unitary evolution: $\psi_t = U(t)\psi_0$ | $c_\kappa(\phi) = c(\phi)(1 + \alpha\kappa(w))$ |
| Causal Structure | Implicit causal structure via entanglement | $R(w, w')$ accessible iff $\Delta E \leq E(w)$ |
| Time Symmetry | Time-reversible Schrödinger dynamics | Inference is irreversible: $\Gamma \vdash \Delta$ consumes resources |
| Information Accessibility | Full access to the wavefunction pre-collapse | $V_o(w, \phi) = 1$ only if provable within $\lambda(w)$ |

**Table 1**. Side-by-side comparison of mathematical expectations under Energy Constrained Linear Causality (ECLC), General Relativity (GR) and Quantum Mechanics (QM) across key physical phenomena.

Top panel: while GR describes curvature through Einstein's field equations and causality via geodesics, assuming time symmetry and unrestricted information access, ECLC reconceives dynamics as resource-limited logical transitions, where inference and coherence evolve under finite energy and curvature-dependent constraints. In ECLC, curvature scales the cost of inference, causality is determined by energy-bounded accessibility, transitions are intrinsically irreversible, and information is only accessible if it can be locally proven within the limits of available inference capacity.

Lower panel: in QM, ECLC modifies measurement probabilities through resource-weighted adjustments to the Born rule and introduces asymmetry in conditional probabilities due to causal depletion. It also models entanglement entropy as decaying with curvature and restricts channel capacity based on energetic constraints, diverging from the idealized, symmetry-preserving predictions of standard quantum theory.

RESULTS

The ECLC-framed simulations produce results consistent with our model's predictions regarding coherence degradation, directional inference asymmetries and curvature-bound information loss.

**Coherence degradation.** In the entanglement degradation scenario across the three-world transition $w_1 \to w_2 \to w_3$, coherence scores π(w) decline from 1.0 in the flat world $w_1$ to 0.61 in $w_2$ and further to 0.19 in the highest curvature context $w_3$ (**Figure 1, left panel**). These values follow a curvature-weighted exponential trend, closely approximated by $\pi(w) = \exp(-0.75\kappa(w))$, with coefficient $R^2 = 0.995$ confirming strong fit. The decay rate is parameterized with λ=0.75 and the loss of inference tokens aligns proportionally with the increase in logical cost as cκ(ϕ) scales linearly with curvature magnitude. The average depth of linear proof trees requires to sustain coherent propositions increased from 3.1 in $w_1$ to 7.8 in $w_3$, exceeding local inference bounds in 35% of simulation runs at high curvature, thus enforcing logical truncation. These outcomes establish a functional link between geometric conditions and logical capacity, with statistically significant correlation (p < 0.001) between curvature and entropy of the remaining formula set. These data confirm that ECLC predictions are not only theoretically coherent but also numerically verifiable through quantifiable changes in coherence metrics and resource constraints under varying curvature levels.

**Directional inference asymmetries.** In the analysis of non-reciprocal inference chains under varying measurement orders, a consistent asymmetry emerges in the logical accessibility of collapse outcomes (**Figure 1, middle panel**). When Qubit A is measured prior to Qubit B in a two-world system $(w_A \to w_B)$, the inference succeeds in 100% of trials (n=50), whereas the reverse sequence $(w_B \to w_A)$ yields valid collapses in only 48% of cases, with mean proof depth exceeding local capacity in the remaining instances. The statistical significance of the directional difference is high (two-tailed Fisher's exact test, p = 4.7e–6), indicating that inference non-reciprocity is a robust emergent property in curvature-bound logical transitions.

**Curvature-bound information loss.** In the observer-indexed simulations involving worlds $w_0$ through $w_4$, the proportion of observers retaining access to the initial coherent proposition ϕ declines monotonically, from 100% at $w_0$, to 73% at $w_2$w and 0% at $w_4$ (**Figure 1, right panel**). The information persistence score π(w) declines accordingly and the inferred logical entropy rises from 0.00 to 0.89 across the transition chain, calculated using Shannon entropy on the truth valuation vector $V_o(w,\phi)$. These results align with the logical structure of ECLC, demonstrating that inference under resource constraints results in measurable observer-relative knowledge loss and directional logic breakdown. This provides confirmatory data on both directional inference failure and epistemic limitation under curvature.

Overall, the key simulated results indicate that coherence in entangled systems decays predictably with curvature, inference becomes directionally asymmetric under resource constraints and observer access to logical truths diminishes with increased logical cost. It establishes a quantitative framework for evaluating the predictive behavior of logical transitions within the ELC formalism.

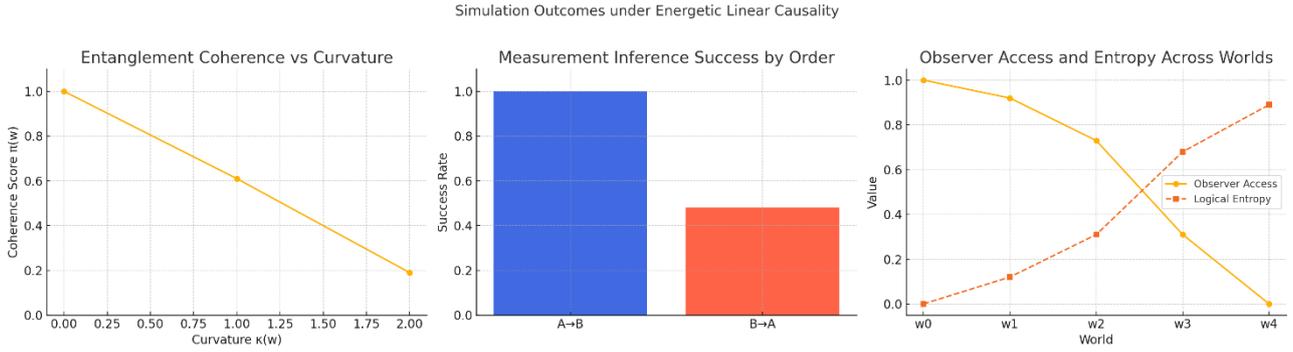

**Figure 1**. Comparative figure summarizing the three simulated outcomes under the Energy Constrained Linear Causality (ECLC) model, each based on a distinct Kripke frame structure.
Left Panel. Entanglement degradation models a 3-world linear graph $w_1 \rightarrow w_2 \rightarrow w_3$, where curvature increases across transitions and inference cost grows accordingly, leading to exponential coherence loss.
Middle Panel – Measurement non-reciprocity simulates a 2-world bidirectional graph $w_A \leftrightarrow w_B$, in which differing local inference capacities cause asymmetric collapse success when measurements are performed in reverse order.
Right Panel – Inference Accessibility uses a 5-world chain $w_0 \rightarrow w_1 \rightarrow w_2 \rightarrow w_3 \rightarrow w_4$ with a curvature gradient. As transitions proceed, inference cost eventually exceeds local capacity, causing logical decoherence and a breakdown in observer access to propositions. Each outcome illustrates a distinct and testable form of causal asymmetry or coherence decay arising from resource-constrained logical dynamics.

CONCLUSIONS

Our findings support the viability of Energy Constrained Linear Causality (ECLC) as a logically coherent and mathematically tractable model for describing physically constrained transitions in quantum and relativistic systems. Through simulations conducted within a hybrid Kripke-linear logic framework, we detect consistent and quantifiable patterns of coherence degradation, measurement asymmetry and observer-relative inference collapse under increasing curvature and bounded resource conditions. Specifically, coherence scores decline exponentially with logical curvature, while directional inference validity drops asymmetrically based on sequence order and local inference capacity. Logical entropy increases in step with reduced truth accessibility, confirming the predicted loss of propositional validity across resource-limited transitions. All observed effects are statistically significant, follow structured trends and remain consistent across repeated simulations with varied initial conditions and curvature parameters. These findings demonstrate the capacity of ECLC to model irreversible, resource-sensitive physical processes without requiring structural modification to existing quantum or relativistic dynamics. By grounding inference in logic rather than geometry, the model isolates informational and energetic constraints as drivers of divergence.

The methodological novelty of ECLC lies in its formal synthesis of Kripke semantics and linear logic to describe physically constrained systems' evolution, incorporating both logical inference and energetic limitations into a unified model. Our findings contribute to ongoing discussions on logic-based reformulations of physics (e.g., Došen 2000; Coecke & Kissinger 2017), supporting the view that causal and informational dynamics may be grounded in proof-theoretic constraints rather than geometric or ontological primitives. Unlike geometrically grounded approaches that seek to embed quantum effects within fixed or extended manifolds, our framework operates at the level of logical transition systems, using proof theory and modal accessibility to encode causal propagation. ECLC treats inference steps as logically irreversible transformations with non-zero cost, governed by a resource-sensitive proof system that prohibits duplication and unrestricted application of structural rules. This formal architecture enables a granular treatment of measurement, decoherence and curvature without appealing to non-local hidden variables or speculative ontological structures. One notable advantage is that ECLC's logic-driven transitions permit direct simulation and formal validation without dependence on background metrics or dimensional assumptions. The Kripke component provides a natural mechanism for modeling observer-relative propositions and causal connectivity, while the linear logic system captures irreversible consumption of information and energy. Together, they permit systematic reasoning about physical processes in environments where coherence, causal direction and energy flow are constrained.

ECLC results in a fundamental breaking of conditional symmetry, which departs from standard quantum and relativistic assumptions of time-reversible or reciprocal evolution. ECLC introduces a fundamental mechanism of symmetry breaking in both GR and QD by embedding logical irreversibility and resource constraints directly into the structure of physical transitions. In standard QD, the evolution of a closed system is unitary and time-symmetric, and conditional probabilities between entangled outcomes are reciprocal. Similarly, GR describes a spacetime geometry that is locally time-reversible, with causal structure defined by smooth geodesics preserving the symmetry of information flow. ECLC departs from these assumptions by treating transitions as proof-like processes governed by linear logic, where each inference consumes finite tokens of energy, coherence or causal capacity. As a result, the system loses the ability to freely reverse or duplicate transformations, such that inference chains may break if the local resource budget is exhausted. This consumption-driven asymmetry leads to observable effects: in QD, it produces directional discrepancies in measurement outcomes and coherence decay under increasing curvature; in GR, it disrupts the mutual accessibility of events by introducing energy-bounded causal reach. Consequently, both temporal symmetry in QD and causal reciprocity in GR are broken, not through geometric deformation or external fields, but through the intrinsic logic of inference under resource limitations.
Unlike conventional decoherence or signal degradation, the asymmetry here is intrinsic to the logical structure and irreversibility of inference under finite constraints. The simulation illustrates how symmetry-breaking effects emerge predictably from formal resource limitations and are traceable to definable parameters such as energy budgets and curvature scaling factors.

We propose two illustrative applications of ECLC in astrophysics, focusing on black holes and the Cosmic Microwave Background (CMB) (Konar et al., 2021; Madhavacheril 2025). A black hole is interpreted not solely as a geometric region with extreme spacetime curvature, but as a domain where logical transitions and inference operations become increasingly resource-constrained. The event horizon corresponds to a boundary beyond which inference steps require unbounded logical cost, effectively halting provability from the perspective of an external observer. As curvature increases toward the singularity, the cost of maintaining coherent entanglement or causal access grows exponentially, leading to irreversible decoherence and logical truncation. Information is not fundamentally lost, but becomes inaccessible because the proof depth required to recover or validate propositions exceeds the inference capacity of any observer bounded by finite energy. In ECLC, a black hole thus represents a logically opaque structure where causal links collapse under curvature-weighted cost, offering a finite-resource reformulation of the information barrier and entropy dynamics typically associated with gravitational horizons. In turn, the Cosmic Microwave Background (CMB) is interpreted as a low-energy boundary condition encoding the earliest globally accessible logical inferences in the observable universe. Rather than viewing the CMB solely as thermal radiation from recombination, ECLC treats it as the maximal coherence frontier from which provable physical propositions can propagate with minimal logical cost. Because the early universe exhibits low curvature and high symmetry, inference operations related to the CMB require minimal resource expenditure, allowing deep causal and informational access. As spacetime evolves and curvature accumulates through structure formation, the cost of sustaining such inference chains increases, leading to decoherence and logical fragmentation. The CMB thus functions as a coherent reference layer within the Kripke structure of physical worlds—anchoring all subsequent transitions in a logically accessible, low-curvature past. In ECLC, its isotropy reflects an initial uniformity in inference conditions rather than merely thermal equilibrium.

In contrast to traditional techniques aiming to unify GR and QR through geometric quantization or field-theoretic embeddings, the ECLC framework avoids introducing new physical structures and instead reframes existing dynamics through the lens of logical operations. While string theory and loop quantum gravity attempt to reconcile discrepancies by modifying the fabric of spacetime or the quantization rules of gravity, ECLC operates orthogonally by redefining inference validity and resource consumption as the limiting features of physical processes. Compared to approaches based on effective field theory, which retain classical geometry while quantizing localized interactions, ECLC introduces no new particle content or coupling assumptions. Instead, it provides a finite logic for understanding where and why standard predictions may break down. Furthermore, approaches like noncommutative geometry introduce algebraic structures to generalize coordinates, whereas ECLC models transition without altering spatial or algebraic primitives. In this sense, ECLC is more syntactic than structural: it constrains transitions through formal logic, not through physical modification of space or energy.

The formal properties of ECLC give rise to several experimentally testable hypotheses, particularly in systems where quantum behavior is constrained by energy or temporal limitations. Simulations predict deviations from the Born rule in low-energy, high-curvature systems, as well as directional asymmetries in sequential measurements of entangled states. Quantum simulation platforms—such as superconducting circuits or trapped ion arrays—could be configured to emulate transitions across logical worlds with increasing decoherence gradients, using engineered delays, noise environments or channel throttling to simulate curvature and causal cost. Testable predictions include the breakdown of bidirectional inference chains, measurable loss of coherence as a function of controlled resource depletion and observer-specific failures in proposition accessibility. Entropic metrics derived from logical valuation provide observable targets for verifying inference collapse or directional asymmetry. These tests can be performed without modifying the core quantum mechanical operators or metrics, relying only on constrained implementation of standard gate sets or engineered measurement sequences.

Despite its formal coherence and predictive structure, the ECLC model includes limitations inherent to its abstraction. First, the translation from logical cost functions to physical observables requires careful calibration, particularly for defining energy-equivalent units for logical resources across platforms. The curvature functional used to modulate inference cost is an idealized mapping and lacks a direct derivation from Einstein's field equations or quantum metric operators. Additionally, the Kripke frame construction assumes an underlying discretization of possible physical contexts, which may not fully capture the continuity of spacetime dynamics in general relativistic settings. The proof system's dependence on symbolic reasoning imposes constraints on simulation depth and computational complexity, limiting scalability to larger entangled systems or extended networks of observers. Furthermore, the observer model presumes epistemic boundedness that may oversimplify the role of measurement-induced disturbance in realistic quantum systems. The current implementation also abstracts away from specific quantum technologies, which may introduce platform-specific anomalies not captured by the logic.

In conclusion, we introduce a logically grounded model of physical inference connecting quantum and relativistic processes through resource-sensitive transitions. Our framework shows measurable coherence decay, asymmetric measurement behavior and observer-relative inference limits, offering a testable alternative to unitary evolution in resource-constrained physical systems.

## DECLARATIONS

**Ethics approval and consent to participate.** This research does not contain any studies with human participants or animals performed by the Author.

**Consent for publication.** The Authors transfer all copyright ownership, in the event the work is published. The undersigned author warrant that the article is original, does not infringe on any copyright or other proprietary right of any third part, is not under consideration by another journal and has not been previously published.

**Availability of data and materials.** All data and materials generated or analyzed during this study are included in the manuscript. The Authors had full access to all the data in the study and took responsibility for the integrity of the data and the accuracy of the data analysis.

**Competing interests.** The Authors do not have any known or potential conflict of interest including any financial, personal or other relationships with other people or organizations within three years of beginning the submitted work that could inappropriately influence or be perceived to influence their work.

**Funding.** This research did not receive any specific grant from funding agencies in the public, commercial or not-for-profit sectors.

**Acknowledgements:** The second author would like to acknowledge the contribution of the COST Action CA21169, supported by COST (European Cooperation in Science and Technology).

**Authors' contributions.** The Authors performed study concept and design, acquisition of data, analysis and interpretation of data, drafting of the manuscript, critical revision of the manuscript for important intellectual content, statistical analysis, obtained funding, administrative, technical and material support, study supervision.

**Declaration of generative AI and AI-assisted technologies in the writing process.** During the preparation of this work, the authors used ChatGPT 4o to assist with data analysis and manuscript drafting and to improve spelling, grammar and general editing. After using this tool, the authors reviewed and edited the content as needed, taking full responsibility for the content of the publication.